# Low-mass star formation in CG1: a diffraction limited search for pre-main sequence stars next to NX Pup *

Wolfgang Brandner[1,2], Jerome Bouvier[3], Eva K. Grebel[4], Eric Tessier[5], Dolf de Winter[6], and Jean-Luc Beuzit[7]

[1] Astronomisches Institut der Universität Würzburg, Am Hubland, D-97074 Würzburg, Germany, brandner@astro.uni-wuerzburg.de
[2] European Southern Observatory, Casilla 19001, Santiago 19, Chile,
[3] Laboratoire d'Astrophysique, Observatoire de Grenoble, Université J. Fourier, B.P. 53, F-38041 Grenoble Cedex 9, France, bouvier@gag.observ-gr.fr
[4] Sternwarte der Universität Bonn, Auf dem Hügel 71, D-53121 Bonn, Germany, grebel@astro.uni-bonn.de
[5] Royal Greenwich Observatory, Madingley Road, Cambridge CB3 0EZ, England, tessier@mail.ast.cam.ac.uk
[6] Astronomical Institute "Anton Pannekoek", University of Amsterdam, Kruislaan 403, NL-1098 SJ Amsterdam, The Netherlands, DOLF@astro.uva.nl
[7] Observatoire de Paris, DESPA (URA 264/CNRS), 5 place Jules Janssen, F-92195 Meudon, France, jlbeuzit@hplyot.obspm.circe.fr

**Abstract.** Using adaptive optics at the ESO 3.6m telescope, we obtained diffraction limited JHK-images of the region around the Herbig AeBe star NX Pup. We clearly resolved the close companion (sep. $0''.128$) to NX Pup – originally discovered by HST – and measured its JHK magnitudes. A third object at a separation of $7''.0$ from NX Pup was identified as a classical T Tauri star so that NX Pup may in fact form a hierarchical triple system. We discuss the evolutionary status of these stars and derive estimates for their spectral types, luminosities, masses and ages.

## 1. Introduction

Cometary Globules (CGs) are cool clouds with compact, dusty, opaque heads, and faint tails stretching away from the head (e.g., Harju et al. 1990). They were first described and defined as a group by Hawarden and Brand (1976) based on Schmidt plates of the Gum Nebula and NGC 5367. Many CGs are found in the Gum Nebula, a large region of ionized gas at a distance of approximately 450pc. In this nebula the isolated neutral globules clearly contrast with the surrounding hot ionized media (Brand et al. 1983).

The heads of most CGs point to an apparent center in the Gum Nebula, suggesting a common origin of the cometary shape of these globules (Zealey et al. 1982). Different mechanisms have been proposed to explain the origin of the tail like structure of the CGs, such as interactions of dense cloud cores with shock waves from a SN explosion (Brand et al. 1983) or ionization shock fronts from nearby OB stars (Reipurth 1983). IRAS observations show an enhanced star formation rate in the CGs of the Gum Nebula with respect to comparison fields (Bhatt 1993). Star formation may have been induced through shocks triggered by the same event that ionized the Gum Nebula complex. Reipurth and Pettersson (1993) identified eight late-type H$\alpha$ emission stars in association with CG4/CG6/Sa101 and CG13. They argue that star formation has occurred in the Gum Nebula for more than $10^6$ years. A mapping of CG1 in molecular lines has been carried out by Harju et al. (1990). They find that CG1 is not in dynamical equilibrium and in CO shows a high (5km/s) velocity component near its head. They conclude that this is shocked material related to star forming processes.

Located at the edge of CG1 is the Herbig AeBe star NX Pup (Irvine 1975). NX Pup is a variable star (Hoffmeister 1949, Strohmeier et al. 1964). It shows H$\alpha$ in emission (Wackerling 1970) and has a strong UV excess (de Boer 1977) and IR excess (Brand et al. 1983, Reipurth 1983).

Reipurth find an MK type of F0-2 III, a total luminosity of 40-50 $L_\odot$, and an age of $10^6$ years, whereas (on the basis of UV observations from space) de Boer (1977) and Tjin A Djie et al. (1984) assign an MK type of A0-1 III. Tjin A Djie et al. and Thé and Molster (1994) derive a total luminosity between 100 and 160$L_\odot$ and an age of $8 \cdot 10^5$ years for NX Pup. The age of NX Pup agrees very well with the age of CG 1 derived from dynamical studies (e.g., Harju et al. 1990).

Observations with the Fine Guidance Sensor system aboard the Hubble Space Telescope reveal that NX Pup (HIC 35488) is a close visual binary with a separation of 0.″126 (Bernacca et al. 1993). Thus all previous evolutionary interpretations should be reevaluated as the total observed luminosity in fact comes from two stars.

Using the ESO adaptive optics imaging system COME-ON+ (CO+) in combination with the SHARP camera we started a diffraction limited imaging survey of intermediate-mass pre-main sequence (PMS) stars listed in "A new catalogue of members and candidate members of the Herbig Ae/Be stellar group" (Thé et al. 1994) with the aim to search for close IR companions. During the first part of this project we obtained JHK images for the NX Pup system. The CLEANed JHK images clearly resolve the close companion, here referred to as NX Pup B, detected in the optical by the HST.

We also found a third, faint PMS star at a distance of 7.″0 from NX Pup A/B, which will be referred to as NX Pup C in the following.

**Table 1.** Journal of observations

| telescope/instr. | date | filter/wavelength |
|---|---|---|
| ESO 1m/vis. phot. | 1.–3.12.1992 | V |
| SAT (LTPV)/vis. phot. | 12.92–2.93 | y |
| ESO 3.6m/CO+ | 1.1.1994 | J,H,K |
| Danish 1.5m/CCD-camera | 8.1.1994 | U,B,V,R,i,H$\alpha$ |
| ESO 1.5m/B&C | 26.1.1994 | 380-760nm |
| NTT/EMMI | 20.3.1994 | 600-725nm |

## 2. Observations and results

### 2.1. Near-infrared Imaging

JHK images of NX Pup with high spatial resolution were obtained in 1994 Jan 1 at the ESO 3.6m telescope. We used the COME-ON+ adaptive optics system (Beuzit and Rousset, 1994) in combination with the SHARP II (System for High Angular Resolution Pictures, see Hofmann et al. 1992) camera from the Max Planck Institute for Extraterrestrial Physics (MPE). The SHARP camera is equipped with a Rockwell NICMOS-3 array. The image total exposure time on NX Pup was 300s in each JHK filter, and similar exposures were obtained on the nearby sky for sky subtraction. Immediately after NX Pup, we observed a reference point source 10′ away (star no.985 in the HST Guide Star Catalogue) which was later used for image deconvolution. The photometric standard HR 3421 was observed under the same conditions to get absolute JHK photometry.

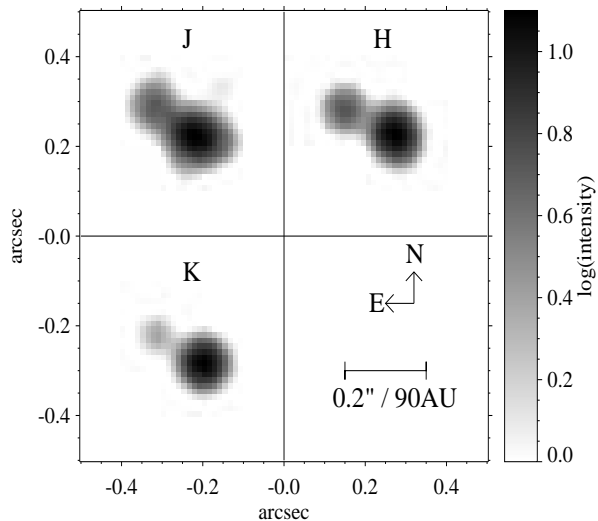

**Fig. 1.** Deconvolved JHK images of NX Pup obtained with the adaptive optics system CO+ at the ESO 3.6m telescope in 1994 Jan 1. The images were rebinned by a factor of 4 (see text for more details). The components A and B (sep. 0.″128, PA 62.4°) are clearly resolved. At a distance of 450pc 0.″128 correspond to a projected separation of 58AU. The faint feature in the J image north-west of NX Pup A is an artefact of the image deconvolution. A logarithmic gray scale was used. Component C is outside these frames. North is up and east is to the left.

The individual exposures in each filter, 5 in J, 150 in H, and 600 in K, were coadded using the "shift and add" (SAA) technique, i.e., coadding the individual images after shifting them so that the centre of the point spread functions coincide.

If a sufficient number ($\geq 50$) of individual exposures is available, one can gain in spatial resolution by selecting images based on a Strehl ratio (SR) criteria. Furthermore, if the single frame exposure times are short, the residual tilt between the individual images is reduced by applying an SAA algorithm. While the latter already gives some improvements mostly in the FWHM, the results of image selection are more significant with respect to the Strehl ratio. However, the rejection rate for image selection must be limited in order to keep a sufficient signal to noise ratio. Table 2 shows the improvement in resolution through SAA and image selection. Out of 600 K band frames 240 were selected and coadded, improving the Strehl ratio from 25% to 31%. Selecting 30 images out of the 150 H band exposures improved the Strehl ratio to 12% – compared to 8% without image selection.

**Table 2.** Gain in spatial resolution (FWHM and Strehl ratio, SR) through shift and add (SAA) and SAA with image selection (IS).

|              | J       | H       | K       |
|--------------|---------|---------|---------|
| FWHM         | 0.″214  | 0.″172  | 0.″156  |
| FWHM (SAA)   | 0.″209  | 0.″157  | 0.″147  |
| FWHM (SAA+IS)| —[a]    | 0.″138  | 0.″143  |
| SR           | 4%      | 8%      | 25%     |
| SR (SAA)     | 4%      | 9%      | 27%     |
| SR (SAA+IS)  | —       | 12%     | 31%     |
| selection rate (IS) | — | 20%     | 40%     |

[a] no IS (only 5 individual exposures)

All these operations were performed with the "local" IRAF package c128 developed by E. Tessier at the Observatory of Grenoble.

On the shift-and-add image in the K band, NX PUP C itself appears to be elongated in the east-west direction (FWHM in east-west direction and north-south direction are 0.″27 × 0.″18, compared to 0.″16 × 0.″16 for the reference source). This elongation is less pronounced in the H band, and is not seen in the J band.

NX Pup C is 7″ away from NX Pup AB, which was used for wave-front sensing. Therefore, we could expect effects from angular anisoplanatism (Wilson and Jenkins 1994). Because of this and the lower signal to noise ratio of NX Pup C, we did not try to deconvolve it.

In the following we will discuss briefly two kinds of anisoplanatism – angular and temporal – which cause image elongations. One effect of angular anisoplanatism is that the Strehl ratio drops and the PSF gets wider with increasing angular distance to the guide star. The FWHM of NX Pup C is systematically wider than the FWHM of the calibration source in each JHK filter. This might be because of angular anisoplanatism. Furthermore, angular anisoplanatism would lead to an elongation of the off-axis PSF in the direction of the guide source (McClure et al. 1991). In the case of NX Pup AB as reference source and NX Pup C as programme star the direction of the elongation would be $\sim 45°$, i.e., from north-east to south-west but not in east-west direction.

Elongations in another direction can be explained by temporal anisoplanatism due to the wind in the dominant turbulent layer (see Roddier et al. 1993 and Wilson and Jenkins 1994). The prevalent wind direction on La Silla near the ground is north-south, but we do not know the wind direction in the dominant turbulent layer during our observations. While an off-axis PSF should suffer from anisoplanatism. However, the on-axis PSF of the reference source does not show any significant deviation from circular symmetry. Summarizing, neither angular nor temporal anisoplanatism are very likely sources of elongation and the elongation may therefore come from the object NX Pup C itself. To know for sure NX Pup C would have to be re-observed.

The resulting coadded images were then deconvolved using a CLEAN algorithm and the nearby reference star as a PSF. The central 0.″5 × 0.″5 part of the CLEANed JHK images is shown in Fig. 1. The close visual companion detected by HST is clearly seen in all filters. This is a clear demonstration of the diffraction limited imaging capabilities of the CO+ adaptive optics system.

Astrometry has been performed on the CLEANed images, whereas photometry of the various components of the system has been performed on the shift-and-add images. We did not carry out the photometry on the CLEANed images as deconvolution algorithms usually do not conserve the flux distribution. Absolute JHK photometry was obtained for NX Pup A+B and for NX Pup C using the IRAF/APPHOT package, while differential photometry between NX Pup A and B was performed within the IRAF/DAOPHOT package using IRAF scripts kindly provided by A. Tokovinine. The JHK magnitudes derived in this manner for NX Pup A, B, and C are listed in Table 3. The overall photometric errors are about 10%.

**Table 3.** JHK photometry (1.1.1994, ESO 3.6m/CO+). For comparison we also show the V magnitudes.

| filter | NX Pup A        | NX Pup B        | NX Pup C          |
|--------|-----------------|-----------------|-------------------|
| J      | $8^m.58$        | $9^m.56$        | $11^m.71$         |
| H      | $7^m.43$        | $8^m.37$        | $10^m.66$         |
| K      | $6^m.15$        | $7^m.90$        | $10^m.10$         |
| V      | $10^m.3^a$      | $10^m.9^a$      | $16^m.03^b$       |

[a] assuming $\Delta V = 0^m.64$ (Bernacca et al. 1993; 1.1.1993) and V=$9^m.8$ for NX Pup A+B
[b] this paper (8.1.1994)

### 2.2. Visual Imaging

CCD images in the Bessel UBVR, Gunn i, and H$\alpha$ filters of the region around NX Pup were obtained in 1994 Jan 8 with the CCD-camera at the Danish 1.5m telescope at La Silla. This camera is equipped with a 1K TEK CCD (ESO #28) and has a scale of 0.″38/pixel.

We obtained two sets of exposures in U,B,V,R,i,H$\alpha$ – short exposures (10s, 5s, 1s, 1s, 1s, 60s) in order not to saturate NX Pup AB and long exposures (180s, 20s, 10s, 10s, 10s, 600s) in order to get reliable photometry of NX Pup C. The seeing was between $1''\!.0$ and $1''\!.5$. The finding chart for NX Pup AB and C is shown in Fig. 2.

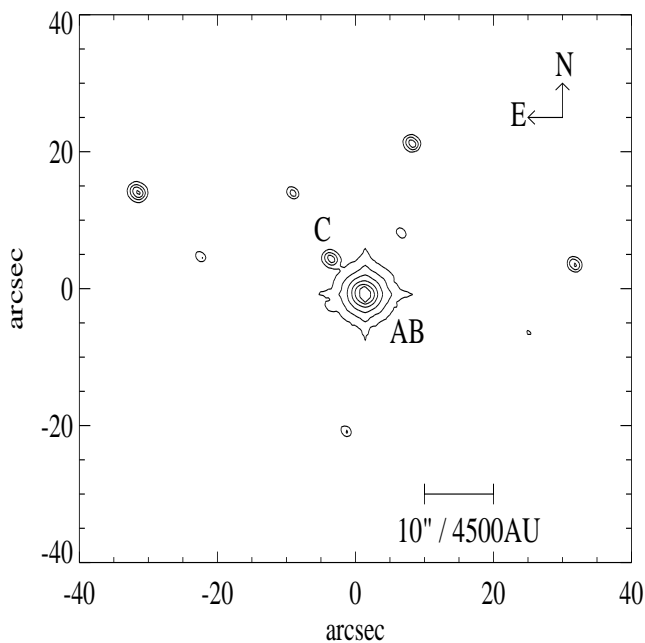

**Fig. 2.** The PMS stars NX Pup A+B (center) and C in a 10s R exposure with the CCD camera at the Danish 1.5m telescope on La Silla. The components AB (unresolved) and C are marked. All other stars within the field show no excess in Hα and therefore are likely to be field stars. North is up and east is to the left.

Two standard fields (SA98 and RU149) of the Landolt list (Landolt 1992) were observed in order to calibrate the photometric data.

The CCD camera at the Danish 1.5m telescope has a shutter delay time of 30ms. For short exposures of 1sec this results in a 3% gradient in intensity from the center to the edge of the field. All short exposures were corrected for the shutter delay. Instrumental magnitudes and transformation equations were computed and solved within the rors are between 2% and 5%. Table 4 lists the results of the visual photometry.

**Table 4.** Visual Photometry (8.1.1994, Danish 1.54m)

| filter | NX Pup AB | NX Pup C |
|---|---|---|
| U | $10^m\!.40$ | $17^m\!.52$ |
| B | $10^m\!.30$ | $17^m\!.46$ |
| V | $9^m\!.78$ | $16^m\!.03$ |
| R | $9^m\!.46$ | $14^m\!.96$ |
| I | $9^m\!.01$ | $13^m\!.68$ |
| R–Hα[a] | $0^m\!.46$ | $0^m\!.68$ |

[a] see definition of Hα magnitude in the text

As NX Pup AB is variable we have to be careful when combining the observations of different epochs.

We do not have an absolute V magnitude for the HST observation of NX Pup, but one of us (DdW) observed NX Pup with the visual photometer at the ESO 1m telescope between 1992 Dec 1 and 3. Furthermore, NX Pup is one of the target stars in the LTPV programme and was observed between 1992 Dec 4 and 22 and 1993 Jan 21 and Feb 8. P.S. Thé was so kind to provide us Strømgren y band measurements of NX Pup – obtained with the Danish 50cm Strømgren Automatic Telescope (SAT) on La Silla, Chile – prior to their publication (Sterken et al. 1995, in prep.). NX Pup was getting fainter from December to February. If we assume the y band measurements to be equivalent to the V magnitudes, simple linear interpolation gives us V = $9^m\!.8 \pm 0^m\!.2$ at the time of the HST observations. Thus NX Pup had about the same V magnitude during the CO+ observations as during the HST observations.

### 2.3. Spectroscopy

We used the ESO 1.5m and the Boller and Chivens (B&C) spectrograph in 1994 Jan 26 to get long slit spectra of NX Pup AB and C simultaneously. The B&C spectrograph is equipped with a 2K FORD chip (ESO #24) and can be rotated as a whole in order to align the slit along a user defined position angle. We used grating #23 with an useful wavelength range from 380 to 760nm and a sampling of 0.19nm/pixel. The FWHM along the slit of the 2D spectra is $4''\!.3$. This relatively large value presumably cannot be attributed to the seeing but to guiding errors during the 30min exposure. Accordingly, the shape of the one dimensional point spread function is better described by a Gaussian than by a function resembling a King or Moffat profile. Because of the large FWHM and the large brightness difference between NX Pup AB and C (5.5 mag in R), the C component is heavily blended.

To extract the spectrum of NX Pup C we fit two Gaussians simultaneously along each scan in spatial direction of our two dimensional spectra. FWHM and separation of the two Gaussians remain fixed, allowing only their heights and the absolute position to vary from scan to scan in the dispersion direction. A comparison to the NTT spectrum (see below) shows good agreement of the spectrum of NX Pup C in the overlapping region, giving us some confidence in our method.

Spectra with a better spectral resolution (0.06nm / pixel) and spatial resolution (FWHM $1''\!.0$) were obtained in 1994 Mar 20 with the ESO Multi-Mode Instrument (EMMI) attached to the NTT. We used the red arm of EMMI, which has a 2K TEK CCD (ESO #36) and a pixel scale of $0''\!.27$/pixel. The exposure time was 15min and the wavelength range 600 to 725nm.

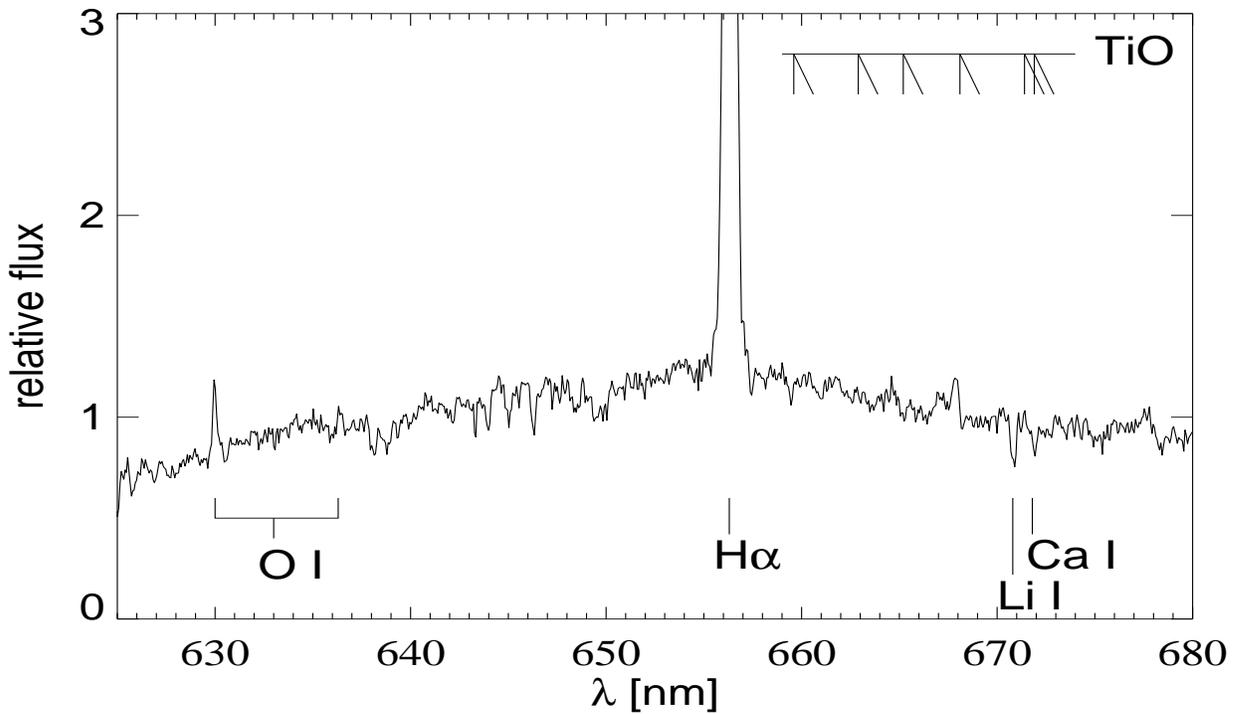

**Fig. 3.** Spectrum of NX Pup C, obtained in a 15min exposure with NTT/EMMI in 1994 Mar 20. Several emission (OI, H$\alpha$) and absorption (TiO bands, Ca I) features are marked. Note the strong Li I 670.8nm absorption, which is a sign of youth.

Because of the large brightness difference between NX Pup C and AB, at least parts of the spectra of NX Pup AB were saturated in both exposures. The saturated regions were excluded from the fits and the final extraction. Two spectrophotometric standard stars were observed in order to derive the sensitivity curve and to achieve a flux calibration. All spectra were extinction corrected applying the default La Silla extinction curve.

Figure 3 shows the medium resolution spectrum of NX Pup C near the H$\alpha$ line. The strong H$\alpha$ emission and the Li absorption can be seen.

## 3. The evolutionary status of NX Pup A+B and C

### 3.1. NX Pup A+B

The position angle (PA) between NX Pup AB and C measured on the CO+ frames and the CCD frames is $45°\!.3\pm0°\!.2$ and the separation $6''\!.98 \pm 0''\!.04$. At a distance of 450 pc this corresponds to a projected separation of 3140AU, so that NX Pup AB and C are unlikely to form a gravitational bound system. On the other hand, this resembles what we would expect to see for a hierarchical triple system: two stars in a close orbit around each other and a third, less massive star in a wide orbit. Only radial velocity measurements and proper motion studies could tell us whether or not the three stars form a physical triple system.

NX Pup A and B themselves are separated by $0''\!.128 \pm 0''\!.008$, as estimated from the JHK images, i.e. 58AU projected separation. They are almost certainly bound, as has been pointed out by Bernacca et al. (1993). The separation and the PA of $62°\!.4 \pm 5°\!.7$ are in good agreement with the values determined by Bernacca et al. ($0''\!.126 \pm 0''\!.007$, $63°\!.4 \pm 1°\!.0$). As there is only a time span of one year between the HST measurements and our CO+ observations, we would not expect signs of orbital motion.

To transform the H$\alpha$ magnitudes, our Bessell R filter can be used as continuum filter. The ratio of our instrumental R and H$\alpha$ filter passbands corresponds to a difference of 3.9 magnitudes (Grebel et al. 1994) in the sense of $R_{inst} \approx H\alpha_{inst} - 3.9$ (for stars without H$\alpha$ emission). A comparison of the ($R_{inst}-H\alpha_{inst} + 3.9$) colours shows that both NX Pup AB and C have a strong excess emission in H$\alpha$ (cf. Table 4). No other stars with R down to 19.3 within a $2'\!.5 \times 2'\!.5$ field centered on NX Pup possess a significant excess in H$\alpha$.

In H-K vs V-K colour-colour diagrams, NX Pup A and B as well as NX Pup C lie outside of the region where reddened main sequence stars with normal extinction would be located. In the J-H vs H-K diagram NX Pup A falls into the region where Herbig AeBe stars can be found. NX Pup B and C lie in the region where reddened main sequence stars and T Tauri stars with low accretion rates and Adams 1992, Strom et al. 1993). This indicates that both NX Pup A and B have intrinsic IR excesses which may arise from a circumstellar disk or envelope. Because of the small separation between A and B, circumstellar material either in form of envelopes or disks should be strongly disturbed. This has already been pointed out by Henning et al. (1994) and could explain the non-detection of NX Pup in their 1.3mm continuum search. The spectral energy distribution (SED) is plotted in Fig. 4. The major part of the system's IR excess emission – already noted by Brand et al. (1983) and Reipurth (1983) and fitted by them with a blackbody of $T_{eff}$ around 1300K – comes from NX Pup A, as evident in our Fig. 4.

As the reddening vector is almost parallel to the sequence of dwarfs and giants with normal colours in the H-K vs V-K diagram, we can constrain the spectral type of NX Pup B. If extinction along the line of sight is negligible, we derive a spectral type earlier than K5 for NX Pup B and earlier than G6 if $A_V=1\overset{m}{.}5$ as for NX Pup A (Hillenbrand et al. 1993). The spectrum obtained with the NTT shows the H$\alpha$ line of NX Pup AB in emission with a central absorption dip. Other lines in emission are the HeI line at 587.5nm, the NaI lines at 589.0 and 589.6nm, OI at 630.0nm and H$\beta$. Because of the strong saturation of the CCD in the central part of the spectrum obtained at the NTT, we did not try to identify photospheric absorption features. The spectrum obtained at the ESO 1.5m shows no evidence for Li I. Martín (1994) has estimated that, e.g., in the combined spectrum of an Herbig AeBe star of spectral type A2 (50L$_\odot$) and an 1.5M$_\odot$ T Tauri type star (10L$_\odot$, $10^6$ yrs) the Li doublet should be visible with an equivalent width of 0.02nm.

**Table 5.** Equivalent width (in nm) of emission (< 0) and absorption (> 0) lines in the spectra of NX Pup AB and C (28.1.1994, ESO 1.5m/B&C; 20.3.1994, NTT/EMMI )

| line | NX Pup AB | NX Pup C |
|---|---|---|
| H$\alpha$ | – [a] | -2.85 |
| OI 630.0 | -0.068 | -0.11 |
| H$\beta$ | 0.25[b] | -0.58 |
| H$\gamma$ | 0.53 | -0.17 |
| Li I 670.8 | – | 0.054 |

[a] double peaked
[b] emission core

The spectral type of NX Pup A has been under debate. Observers in the UV found systematic earlier spectral types (A0–A2, de Boer 1977, Tjin A Djie et al. 1984) than observers in the visual (F0–F2, Brand et al. 1983, Reipurth 1983). Recently, Blondel and Tjin A Djie (1994) suggested that the low resolution IUE spectrum of NX

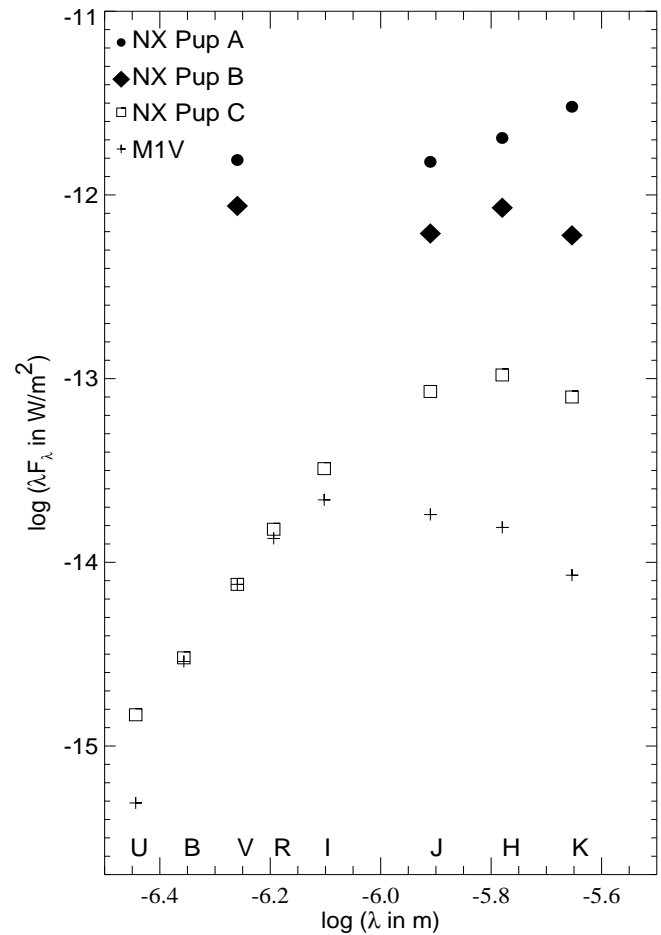

**Fig. 4.** Spectral energy distribution $\lambda F_\lambda$ of NX Pup A and B (VJHK) and C (UBVRIJHK). For comparison we also show the spectral energy distribution of a M1 type star of the same apparent V magnitude as NX Pup C. Note that the SED is rising towards longer wavelength for NX Pup A. The SED of NX Pup B and C peaks near 1.5$\mu$m. NX Pup C show IR excess and UV excess. The errors in flux are 5% or less.

Pup A/B can be decomposed into the spectrum of a hot boundary layer – situated between accretion disk and star – and the photospheric spectrum of an F2 type star.

We use the spectral type – effective temperature calibration from de Jager and Nieuwenhuijzen (1987) to determine $T_{eff}$. If we take the extinction values from Blondel and Tjin A Djie ($A_V = 0\overset{m}{.}4–0\overset{m}{.}7$), assume a spectral type A7–F2 and V=$10\overset{m}{.}1–10\overset{m}{.}5$ for NX Pup A, we get a luminosity of 15–29 L$_\odot$. With the knowledge of the luminosity and the effective temperature we are able to place NX Pup A on an HR diagram and to determine its evolutionary status. New sets of pre-main sequence evolution tracks have been computed by D'Antona and Mazzitelli (1994). From their tracks based on opacities by Alexander et al. (1989) and the convection model from Canuto and Mazzitelli (1990) we get a mass of 2$M_\odot$ and an age around $5 \times 10^6$ yr (see Table 6 and Fig. 5).

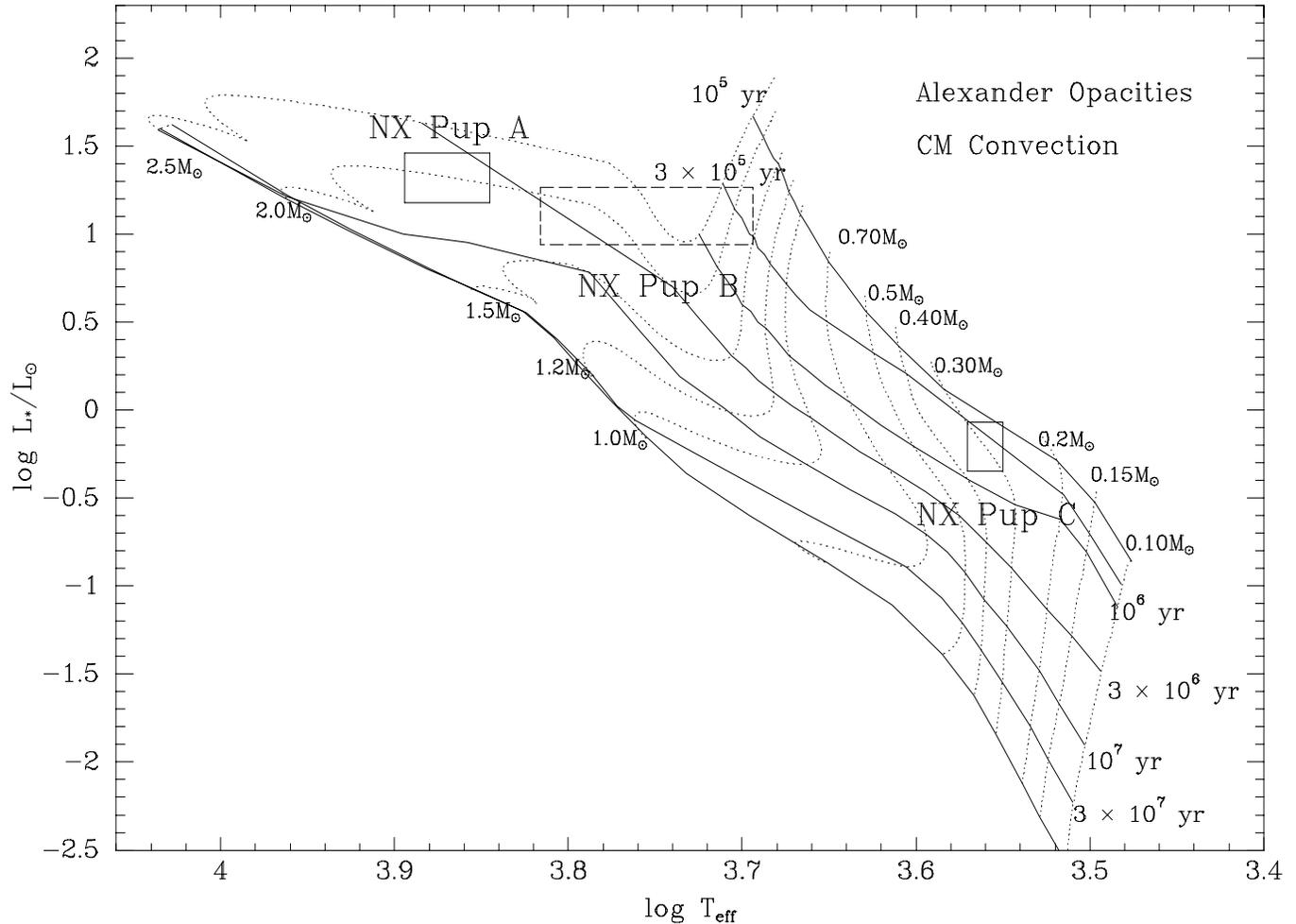

**Fig. 5.** HR-diagram. The pre-main sequence evolutionary tracks are from D'Antona and Mazzitelli (1994). The solid lines mark the isochrones and the zero-age main sequence, the dotted lines are the evolutionary tracks for stars in the mass range from 0.1 to 2.5$M_\odot$. The positions of NX Pup A, B, and C in this diagram are marked by boxes. For NX Pup A we assume a spectral type A7–F2, V=$10^m\!.1$–$10^m\!.5$, and $A_V$= $0^m\!.4$–$0^m\!.7$ (Blondel and Tjin A Djie 1994). For NX Pup B we assume the same extinction, a spectral type F5–G8, and V=$10^m\!.7$–$11^m\!.1$. The evolutionary status of NX Pup C is better defined, yielding an age around $5\times10^5$yrs and a mass of 0.30$M_\odot$.

The evolutionary status of NX Pup B is less well constrained. From its position in the colour–colour–diagrams and the fact, that it is fainter than NX Pup A, a spectral type between mid F and late G seems to be likely. Assuming V=$10^m\!.7$–$11^m\!.1$ and the same extinction values as for NX Pup A, we get a luminosity between 9 and 18 $L_\odot$. The dashed box in Fig. 5 shows that an age from 0.3 to $5 \times 10^6$ yrs and a mass between 1.2 and 2.5$M_\odot$ is possible.

Spatially resolved observations of NX Pup A and B redward of 2.2$\mu$m would be desirable. However, because of the diffraction limit they have to wait for the next generation of (southern) telescopes with their 6 to 8 m mirrors. In the meantime speckle observations in the R and I band and speckle spectroscopy could provide us with better estimates on luminosity and spectral type of NX Pup A and B.

### 3.2. NX Pup C

NX Pup C itself is unresolved on the shift-and-add images. In the following we regard it as a single star.

In the spectra of NX Pup C the Li I 670.8nm absorption line is present with an equivalent width of 0.054nm (Fig. 3). Since Li is depleted in the deeper zones of the convection layer of low-mass stars, the presence of this line is still in its PMS evolution phase.

The equivalent width of the Hα emission is 2.85nm (Table 5), thus NX Pup C belongs to the group of classical T Tauri stars (CTTS). TiO bands at 665.2 and 668.1 nm (typical for M-type stars) are visible, whereas TiO bands at 559.8, 562.9 and 556.1 nm (present from M2 on) are not visible. The spectral A index, which estimates the intensity of the CaH band around 697.5nm (Kirkpatrick et al. 1991), yields a possible MK type between M0.5V and M4III. Taking uncertainties in the extinction correction (see below) into account as well as the fact that NX Pup C as a PMS star is still in its contraction phase, we get a spectral type between M0.5 and M1.5.

In order to derive NX Pup C's luminosity, an estimate of the extinction on the line of sight has to be derived. This is not straightforward, since we have three constituents to the extinction: foreground extinction, extinction within CG1 (radio maps from Harju et al. show that the molecular cloud extends well beyond NX Pup) and possibly circumstellar extinction. Another difficulty when estimating the star's luminosity arises from the presence of both an UV excess (see Table 4) and a near-IR excess (as revealed by the location of NX Pup C in the V-K v. H-K colour-colour diagram, see above). The strong UV excess may be explained by the presence of the Balmer jump, a feature commonly observed among CTTS (e.g. Valenti et al. 1993), while the near-IR excess implies that part of the stellar radiation is reprocessed within the disk. These excesses and the presence of strong Balmer lines indicates ongoing accretion in the disk of NX Pup C. Therefore, the observed total luminosity is the sum of the photospheric luminosity and the accretion luminosity.

As the maximum contribution of the stellar photosphere to the total spectral energy distribution (SED) is at about 1μm (Bertout et al. 1988, Hartigan et al. 1992), the I or J magnitudes are the best measures of the stellar photospheric flux. The presence of photospheric absorption lines tells us that the veiling is not too strong. However, high-resolution spectra would be necessary for an exact evaluation of the veiling.

We use the R–I colour and a spectral type of M1V for NX Pup C to compute the extinction. Colours and bolometric corrections for M dwarfs are from Hartigan et al. (1994). As interstellar extinction relations are colour-dependent we use the new values for differential and total extinction from Grebel and Roberts (1994). Assuming a normal interstellar extinction law towards NX Pup and a distance of 450pc, we derive a total extinction $A_I$=0.93 ($A_V$=1.47) and a luminosity of 0.85$L_\odot$ for NX Pup C. If we use the V–R colour instead of R–I, we get $A_I$=0.43 ($A_V$=0.68) and a luminosity of 0.45$L_\odot$ for NX Pup C. This illustrates the difficulty in deriving extinction values for stars with abnormal spectral energy distributions like classical T Tauri stars.

in Fig. 4 and compared to that of an M1 dwarf. The UV and near-IR excess above the photospheric SED are clearly seen, the latter amounting to $\Delta(K) \approx 0.5$dex, typical for classical T Tauri stars (Strom et al. 1989).

We use again the relation from de Jager and Nieuwenhuijzen (1987) to determine $T_{eff}$ for NX Pup C. In the region between M0.5 and M1.5, $T_{eff}$ is almost independent of the luminosity class. Taking the uncertainties in the spectral classification into account, we get an effective temperature between 3550 and 3720K.

From the tracks of D'Antona and Mazzitelli we get a mass of 0.30$M_\odot$ and an age of $5 \times 10^5$ yrs for NX Pup C (see Table 6 and Fig. 5).

**Table 6.** Evolutionary status of NX Pup A, B, and C

| NX Pup | A | B | C |
|---|---|---|---|
| sep. | – | $0''.128 \pm 0''.008$ | $6''.98 \pm 0''.04$ |
| PA | – | $62°.4 \pm 5°.7$ | $45°.3 \pm 0°.2$ |
| SpT | A7–F2[a] | F5–G8 | M0.5–M1.5 |
| $L/L_\odot$ | 15–29 | 9–18 | 0.45–0.85 |
| age | $5 \times 10^6$ yrs | $0.3$–$5 \times 10^6$ yrs | $5 \times 10^5$ yrs |

[a] Brand et al. 1983, Reipurth 1983, Blondel & Tjin A Djie 1994

Our results are sensitive for the choice of evolutionary tracks. Tracks based on convection models with mixing-length theory yield a somewhat higher age for NX Pup C ($10^6$ yrs).

## 4. Summary

We investigated the properties of stars located close to the Herbig Ae/Be star NX Pup A. We found that its close companion, NX Pup B, is very likely a pre-main sequence star exhibiting strong IR excess, though the lack of a spectrum does not allow us to assess its mass and age. Nevertheless, its somewhat lower luminosity than NX Pup A, and its position in colour-colour-diagrams suggest a spectral type between mid F and late G. The more distant companion, NX Pup C, which may or may not be physically associated to NX Pup A/B, is identified as a low-mass classical T Tauri star whose age around $5 \times 10^5$ years is somewhat younger than the ages deduced by others for NX Pup AB and CG1. This provides the first evidence that low-mass star formation occurred in this cometary globule.

*Acknowledgements.* We thank P.S. Thé and the LTPV program for the permission to use their photometric data on NX Pup prior to publication and A. Tokovinine for providing IRAF scripts he developed to perform photometry of close binaries. We would like to thank Thomas Lehmann for providing his at the Danish 1.5m telescope. Discussions with R. Wilson are gratefully acknowledged.

WB and EKG were supported by student fellowships of the European Southern Observatory.

This research has made use of the Simbad database, operated at CDS, Strasbourg, France, NASA's Astrophysics Data System (ADS), version 4.0, and the IRAF PACKAGE C128, developed by E. Tessier at the Observatory of Grenoble.